\documentclass[preprint]{aastex}
\def\sss{\scriptscriptstyle}
\def\^#1{^{\sss #1}}
\def\_#1{_{\sss #1}}
\def\beq{\begin{equation}}
\def\eeqno#1{\label{#1}\end{equation}}

\def\rarrow{\rightarrow }

\def\anom{a}

\def\dleft{\rlap{{\it D}}\raise 8pt\hbox{$\scriptscriptstyle\Leftarrow$}}
\def\dright{\rlap{{\it
D}}\raise 8pt\hbox{$\scriptscriptstyle\Rightarrow$}}

\def\cos{{\rm cos}}
\def\sin{{\rm sin}}
\def\dmx{\d\m(x)}

\def\kms{{\rm km~s^{-1}}}
\def\cmss{{\rm cm~s^{-2}}}
\def\kpc{{\rm Kpc}}
\def\pc{{\rm pc}}
\def\au{{\rm a.u.}}
\def\aspc{{\rm "/c}}

\def\msun{M\_{\odot}}
\def\az{a\_{0}}

\def\l0{\ell\_{0}}

\def\rar{\rightarrow}
\def\s{\sigma}

\def\l{\lambda}

\def\f{\phi}
\def\t{\theta}
\def\k{\kappa}
\def\eN{\eta\_N}

\def\r{\rho}
\def\rp{\rho_p}
\def\trp{\tilde\rho_p}
\def\hrp{\hat\rho_p}
\def\m{\mu}
\def\tmu{\tilde \m}
\def\bmu{\bar \m}
\def\n{\nu}
\def\tnu{\tilde \n}
\def\bnu{\bar \n}

\def\U{\mathcal{U}}

\def\P{\mathcal{P}}
\def\tP{\tilde\P}
\def\hP{\hat\P}
\def\o{\omega}

\def\D{\Delta}

\def\d{\delta}

\def\drt{d^3\vr}
\def\a{\alpha}
\def\b{\beta}
\def\c{\gamma}
\def\d{\delta}
\def\xlimin{{x\rarrow\infty \atop{\raise 1pt\hbox to 30pt{\rightarrowfill}}}}
\def\limlim#1#2{{#1\rarrow #2 \atop{\raise 1pt\hbox to 30pt{\rightarrowfill}}}}

\def\vr{{\bf r}}
\def\ve{{\bf e}}
\def\vR{{\bf R}}

\def\vF{{\bf F}}
\def\vD{{\bf D}}

\def\vg{{\bf g}}
\def\vgN{{\bf g}\_N}
\def\vu{{\bf u}}

\def\vn{{\bf n}}

\def\va{{\bf a}}

\def\vF{{\bf F}}

\def\tRz{\tilde R_0}
\def\ve{{\bf e}}

\def\S{\Sigma}
\def\vds{{\bf d\sigma}}

\def\grad{\vec\nabla}
\def\div{\vec \nabla\cdot}
\def\gf{\grad\phi}

\begin{document}

\title{MOND effects in the inner solar system}
\author{Mordehai Milgrom }
\affil{ The Weizmann Institute Center for Astrophysics}

\begin{abstract}
I pinpoint a previously unrecognized MOND effect that may act in the
inner solar system, and is due to the galactic acceleration,
$g_g=\eta\az$: a byproduct of the MOND external-field effect.
Predictions of the effect are not generic to the MOND paradigm, but
depend on the particular MOND formulation at hand. However, the
modified-Poisson formulation, on which I concentrate, uniquely
predicts a subtle anomaly that may be detected in planetary and
spacecraft motions (and perhaps in other precision systems, such as
binary pulsars), despite their very high accelerations, and even if
the MOND interpolating function is arbitrarily close to unity at
high accelerations. Near the sun, this anomaly appears as a
quadrupole field, with the acceleration at position $\vu$ from the
sun being $g^{\anom}_i(\vu)=-q^{\anom}_{ij}u^j$, with
$q^{\anom}_{ij}$ diagonal, axisymmetric, and traceless:
$-2q^{\anom}_{xx}=-2q^{\anom}_{yy}=q^{\anom}_{zz}=q(\eta)(\az/R\_M)$,
where $R\_M=(\msun G/\az)^{1/2}\approx 8\cdot 10^3\au$ is the MOND
radius of the sun. The anomaly is described and analyzed as the
Newtonian field of the fictitious cloud of ``phantom matter'' that
hovers around the sun. I find, for the relevant range of $\eta$
values, and for a range of interpolating functions, $\m(x)$, values
of $10^{-2}\lesssim - q\lesssim 0.3$, which turn out to be sensitive
to the form of $\m(x)$ around the MOND-to-Newtonian transition. This
range verges on the present bounds from solar system measurements.
There might thus exist favorable prospects for either measuring the
effect, or constraining the theory and the relevant parameters.
Probing this anomaly may also help distinguish between modified
inertia and modified gravity formulations of MOND. I also discuss
briefly an anomaly that is generic to MOND in all its formulations,
and competes with the quadrupole anomaly in the special case that
$1-\m(x)$ vanishes as $x^{-3/2}$ as $x\rar \infty$.
\end{abstract}


\section{Introduction--the MOND external-field effect}
The MOND external-field effect (EFE) encapsulates the fact that even
a constant-acceleration gravitational field in which a system is
falling may enter and affect the internal dynamics of this system.
It results from the non-linearity of MOND, which, in turn, follows
from the basic premises of the paradigm. This means that even though
all the constituents of the system fall in the same external field,
this field does not cancel in their relative motion, as happens in
(the linear) Newtonian dynamics. The effect appears already in the
pristine formulation of MOND (Milgrom 1983), as well as in the
formulation of MOND as a modified gravity described by a nonlinear
extension of the Poisson equation (Bekenstein \& Milgrom 1984,
Milgrom 1986a). Many applications to astrophysical systems have been
discussed to date, including dwarf spheroidal galaxies in the field
of a mother galaxy (Brada \& Milgrom 2000a, Angus 2008), globular
clusters (Haghi et al. 2009), warp induction by a companion (Brada
\& Milgrom 2000b), escape speed from a galaxy (Famaey, Bruneton, \&
Zhao 2007, Wu et al. 2007), binary galaxies (Tiret et al. 2007),
departure from asymptotic flatness of the rotation curve (Wu et al.
2007), galaxy clusters (Angus \& McGaugh 2008), and others.
\par
In all these applications one deals with accelerations, both
internal ($\vg\_{in}$) and external ($\vg\_{ex}$), that are of order
of, or smaller than, the MOND acceleration constant $\az$, and with
external accelerations that are similar or larger than the internal
ones. In these cases the pristine formulation and the
modified-Poisson equation give similar results, barring
order-of-unity differences in the geometry. In modified-inertia
formulations of MOND the EFE still acts, but its implications can be
quite different (Milgrom, in preparation).
\par
Here I consider a situation in which the relative magnitudes of the
three accelerations are very different:  $g\_{in}\gg g_g>\az$ (in
what follows I use everywhere $\vg_{ex}=\vg_g$ for the galactic
field). This characterizes, for example, high acceleration systems
in the galactic neighborhood of the sun, such as the inner solar
system (SS) or binary stars. These three inequalities would each,
separately, reduce the effect, and render it small. However, the
effect might still be detectable in systems that are measured with
high accuracy such as motions in the solar system, or in binary
pulsars. This arrangement of accelerations does not lend itself to
simple analytic, or numerical, considerations, as do the cases
studied so far.

\par
Small, nonrelativistic MOND anomalies in high-acceleration motions
may possibly have several origins. In non-local, modified inertia
formulations of MOND, anomalies may appear in the inner solar system
in the motion of bodies that are on highly eccentric trajectories;
trajectories that take them to large distances, where accelerations
are low (Milgrom, in preparation). Cases in point are the
long-period comets, and the Pioneer spacecraft. Because of the
non-locality of such theories, the motions in high-acceleration
portions of the orbit may be affected by the low acceleration ones,
giving rise to small anomalies. Such MOND effects have been proposed
as a possible mechanism for generating the Pioneer anomaly, without
affecting the motions of planets, whose orbits are wholly in the
high acceleration regime (Milgrom 2002, 2005, in preparation).
\par
I concentrate on nonrelativistic, modified-gravity formulations, in
which MOND is described by a modified potential field, in which
particles move according to the standard laws of motion. In this
case, an anomaly can arise in high acceleration systems due to a
remaining departure of the MOND interpolating function, $\m(x)$,
from unity, even at high accelerations.\footnote{Galactic dynamics
probe this function only at low values of $x\lesssim 5$, and gives
no constraint whatever on the high $x$ behavior. It is easy to
invent forms of $\m(x)$ with slow rise near $x=1$, but very fast
convergence to $1$ at high $x$ (see some examples in Milgrom \&
Sanders 2008).}$^,$\footnote{Anomalies may also appear in
relativistic theories such as TeVeS and the like, because these
theories do not exactly coincide with general relativity even in the
limit $\az\rar 0$ (see Bekenstein 2006, Sanders 2006, 2007, Sagi
2009, and Skordis 2009.)} For example, in the inner SS such a
departure appears  as an anomalous, spherical acceleration field.
Its effects on planetary motions, and constraints on the high-$x$
behavior of $\dmx\equiv 1-\m(x)$, have been discussed, e.g., in
Milgrom (1983), and in Sereno \& Jetzer (2006). In such analyses the
sun is treated as an isolated point mass.
\par
Here, I discuss another type of anomaly having to do with possible
influence of the galactic field through the EFE. This, as we shall
see, appears in the inner SS as an anomalous quadrupole field, not
as a spherical one.
\par
In order to isolate this effect from the first type, I shall use,
all along, forms of $\m$ with $\dmx\rar 0$ fast enough to dwarf the
first effect in comparison with the EFE.
\par
It is worth noting first that for the parameter values relevant to
the SS, there is no unique MOND prediction of the EFE. The
primitive, pristine formulation (Milgrom 1983), in which the MOND
acceleration, $\vg$, is calculated from the local Newtonian value,
$\vgN$, via an algebraic relation of the form
 \beq \vg=\n(|\vgN|/\az)\vgN \eeqno{algebraic} [where $y\n(y)=x$ is
the inverse of $x\m(x)=y$], captures the main MOND effects in many
instances, for example, in describing galaxy rotation curves, or the
EFE when $g_g\gg g\_{in}$ (Milgrom 1983). However, it is quite
inapplicable to the case $g_g\ll g\_{in}$, such as in solar system,
as it is known not to give correctly the motion of such compound
systems in an external field: For example, atoms in a star, or the
stars in a close binary, moving in the outskirts of a galaxy are
subject to high total accelerations because of the internal forces
in the system. The naive formulation will say then that the full
motion of these constituents is Newtonian. But this would imply that
the whole star (or binary) is also moving Newtonically in the
galaxy, contrary to what is observed, and to what we want MOND to
accomplish. This is also the situation for the solar system, as the
acceleration of the Sun due to Jupiter alone is $\sim 2500\az$, and
that of all the planets due to the sun, even larger. Applying the
algebraic formulation to the motions of the Sun and planets,
separately, would give not only the wrong motion for the solar
system in the Galaxy, but also a negligible EFE for a fast
decreasing $\dmx$.
\par
Modified-inertia formulations of MOND are more versatile and it is
not possible to deduce a generic modified inertia prediction of the
EFE (Milgrom, in preparation). For example, such formulations can be
constructed in which there is no EFE in the inner SS. This is the
case, e.g., in the toy theory described in Milgrom (1999). (This
theory does exhibit an EFE in other circumstances.) But we do not
yet have an acceptable formulation of MOND as modified inertia.
\par
However, it turns out that in the modified Poisson formulation of
Bekenstein \& Milgrom (1984) there is an EFE that is felt in the
inner SS as a quadrupole anomaly (to dominant order). This anomaly
exists despite the high accelerations of the planets, and,
interestingly, it remains finite even with $\dmx\rar 0$ arbitrarily
fast at high $x$. Furthermore, the strength of the effect is
sensitive to the form of $\m(x)$ in the MOND-to-Newtonian transition
region $x\sim 1$, even though the effect itself is probed at very
high accelerations.\footnote{In Milgrom (1986a) I discussed the
anomalous potential difference that is predicted between the sun's
vicinity and infinity. This, however, does not affect motions near
the sun, and depends on system parameters, and on the choice of
$\m(x)$, very differently from the quadrupole anomaly discussed
here. For example, for the isolated case, $\eta=0$, the quadrupole
vanishes, while the potential anomaly becomes infinite.}
\par
All this results from properties of the Poisson equation, and its
ilk, such as the non-linear version treated here: Distortion of the
field in one place is felt in all other regions. Here, the
distortion is created in the sub $\az$ region that must always exist
in the direction of the galactic center, due to the tag-of-war
between the Sun and the Galaxy.
\par
My main purposes here are to point out the anomaly, describe its
nature,  calculate its strength, and describe some of its effects on
planetary motions. I leave more detailed and systematic study to
future work.
\par
In section \ref{anomaly}, I describe the anomaly in the frame of the
modified Poisson theory. In section \ref{multi}, I briefly comment
on multi-potential theories, such as the nonrelativistic limit of
TeVeS. In sections \ref{surrogate}, I describe an approximation for
computing the effect. In section \ref{results}, I give the numerical
results for various forms of $\m(x)$ both for numerical solutions of
the full field equation and for the approximate expression. In
section \ref{solar}, I look at solar system effects, mainly the
induced anomalous perihelion precession. Section \ref{discussion} is
a discussion. The appendices describe the properties of the
fictitious ``phantom matter'' that might be viewed as giving rise to
the anomaly within a Newtonian framework.

\section{The quadrupole anomaly in the modified Poisson theory }
\label{anomaly} The special case of the EFE studied here affects all
high-acceleration galactic systems and could, in principle, be
detected in precision systems such as the SS, binary pulsars, etc..
I shall concentrate here on the EFE for a point mass, because it is
simpler to treat, involving, as it does, only $MG$, $\az$, and
$\eta=g_g/\az$ as defining parameters. While the principles are
similar, the treatment of a general few-body system is more
complicated, because it involves several additional dimensionless
parameters, such as characteristic lengths in units of the MOND radius of the system, $R\_M=(M
G/\az)^{1/2}$, the mass ratios, orientation with respect to the
external field, etc.. Even the isolated, two-body problem is not
exactly solvable in MOND, while the test particle motion in the
field of a point mass is. The EFE for few-body systems will be
discussed elsewhere.
\par
Accordingly, I consider here the motion of a test particle (e.g., a
planet) in the field of a point mass (e.g., the sun), embedded in an
asymptotically constant gravitational field. The latter
approximates, e.g., the galactic field at the position of the
sun.\footnote{In principle, we have to solve for the combined
Galaxy-sun field. Our approximation amounts to taking the Galaxy as
a mass $M\rar\infty$, at a distance $R\rar\infty$, with $M/R^2$
fixed. Corrections to this approximation are expected to be of order
$(\msun/M)^{1/2}\sim 3\cdot 10^{-6}$.} In this case, the
acceleration of the test particle is given by the gradient of the
modified potential field (not so for a non-test mass); so, we set to
calculate the anomalous potential field for this problem.
\par
In the modified Poisson theory of Bekenstein \& Milgrom (1984), the
gravitational potential, $\f(\vr)$, due to a density $\r(\vr$),
embedded (falling) in a constant external field $\vg_g$, is gotten
from the nonlinear generalization of the Poisson equation:
 \beq \div[\m(|\gf|/\az)\gf]=4\pi G\r.  \eeqno{poisson}
Here $\m(x)$ is the interpolating function characterizing the
theory. This equation has to be solved with the boundary conditions:
$\gf\rar -\vg_g$. The internal dynamics is then governed by the
field $\grad\f\_{in}=\gf+\vg_g$. The MOND anomaly is that part of
$\gf\_{in}$ in excess of the Newtonian field for the same mass when
isolated, call it $-\gf\^N$.  The MOND anomaly is then
  \beq \vg^{\anom}=-\gf_a=-(\gf\_{in}-\gf\^N). \eeqno{diffa}
We want to determine the behavior of this anomaly at distances from
the Sun much smaller than $R\_M$, where the accelerations are much
higher than $\az$.
\par
The above anomalous acceleration describes all MOND effects
combined. When $\dmx=1-\m(x)$  vanishes at high $x$ as $A x^{-\a}$,
we get an anomaly near the sun, even in the absence of an external
field. This isolated-mass acceleration is spherical and pointing
towards the sun,  $\vg^{\anom}=-A\az(u/R\_M)^{2(\a-1)}\ve$, where
$\ve=\vu/u$ is the radial unit vector and $\vu$ is the position with
respect to the Sun (Milgrom 1983). For example, for $\a=1$ we get a
constant anomalous acceleration pointing towards the sun
$\vg^{\anom}=-A\az\ve$. This could explain the Pioneer anomaly, but
produces too strong effects on the planets (as already shown by the
analysis of Milgrom 1983). In a recent detailed analysis, Sereno \&
Jetzer (2006) conclude that SS measurements roughly allow only
$\a\gtrsim 1.5$ (depending on the value of $A$).
\par
Sereno \& Jetzer take the parameter values (such as perihelion
precession rate, or semi-major axis) quoted as yielding the minimum
best ephemerides fit residuals, as actual limits on any possible
anomalous force, not included in the fits. They thus obtain limits
on the strength of such various forces, including MOND anomalies for
various $\a$. However, one may question this procedure, since the
more sensible approach would be to include any suspected anomalous
force in a global fit, and in this way to determine what strength of
the anomaly can be tolerated by the data (I am grateful to Gilles
Esposito-Farese for making this point to me). For example, using the
above criterion, Sereno and Jetzer deduce an upper limit of
$10^{-10}\cmss$ on a constant, anomalous, sunward acceleration that
can be tolerated by the data for Uranus, $4\cdot 10^{-10}\cmss$ for
Neptune, and $10^{-9}\cmss$ for Pluto. However, Fienga et al.
(2009), following the more sensible procedure, and including a
constant acceleration in the equations of motion of these three
planets in their global fitting procedure (to test the tolerance to
a Pioneer-like anomaly), find that the Uranus data can tolerate a
constant acceleration of up to $\approx 2\cdot 10^{-8}\cmss$, and
that of Neptune and Pluto allow an anomalous constant acceleration
even of the size of the claimed Pioneer anomaly; namely $\approx
8\cdot 10^{-8}\cmss$. These are some two order of magnitude larger
than the limits deduced by Sereno and Jetzer (2006). Clearly, the
whole question of SS constraints on the high acceleration behavior
of $\m(x)$ requires revisitation.
\par
Be that as it may, the case $\a=3/2$ happens to be interesting
conceptually: It gives rise to a harmonic, anomalous acceleration
$\vg^{\anom}=-A\az\vu/R\_M$, which has the same $u$ dependence, and a
similar magnitude, as the quadrupole anomaly that is my main subject
in this paper.
\par
In most of what what follows I isolate the EFE from such spherical
anomalies by assuming forms of $\m$ with fast decreasing $\dmx$ at
large values of $x$. I will, however, discuss, in section
\ref{solar}, the effects of the harmonic anomaly produced in the
case of the critical exponent $\a=3/2$, which is interesting in the
present context. It is an effect that is generic to MOND and obtain
in any formulation as long as the characteristic interpolation
function has the appropriate asymptotic form. The two effects are
separate and largely additive, since the quadrupole anomaly depends
mainly on the behavior of $\m$ near the transition region, while the
harmonic anomaly derives from the high $x$ behavior of $\m(x)$.
\par
What then is $\vg^{\anom}$ like near the Sun under the above
restriction on $\m$? Note first, importantly, that $\vg^{\anom}$
vanishes identically at the position of the Sun (shown in Appendix
\ref{A1}). Secondly, note that with our choice of $\m(x)$, the
anomaly is practically harmonic (a solution of the Laplace equation)
within a large volume around the sun. The reason for this is as
follows: in this region $|\gf|/\az\gg 1$ and so, practically, $\m=
1$, there. As a result, $\f$ solves the (linear) Poisson equation
there, with the same density source (the sun) as $\f\^N$, and hence
their difference is harmonic. (For $\a\le3/2$ this is not the case:
for Example, for $\a=3/2$ we have a constant phantom density near
the sun.) Add to these the fact that the only length scale in the
problem is $R\_M\approx 8\cdot 10^3\au$, and that the dimensionless
parameter $\eta$ is of order unity, and we conclude that when
$|\vu|$ is small enough with respect to $R\_M$, $\f_a(\vu)$ is
dominated by a quadrupole (a dipole is absent because the anomalous
acceleration vanishes at $\vu=0$); so we can write there
 \beq\f_a(\vu)\approx \f_a(0)+{1 \over 2}q^{\anom}_{ij}u^iu^j, \eeqno{kuiya}
where the anomalous quadrupole is diagonal, and axisymmetric in a
frame where the $z$ axis is along the external field. Harmonicity of $\f_a$ implies that $q^{\anom}_{ij}$ is traceless:
 \beq q^{\anom}_{xx}=q^{\anom}_{yy}=-q^{\anom}_{zz}/2. \eeqno{gugug}
 The effect is thus characterized by
one quantity that we have to determine, say $q^{\anom}_{zz}$. On
dimensional grounds we can write:
 \beq q^{\anom}_{zz}=q(\eta)\az R\_M^{-1}, \eeqno{nuter}
 and I seek to determine the
dimensionless, quadrupole parameter $ q(\eta)$, which describes the
EFE anomaly.\footnote{Since the EFE anomalous acceleration is linear
in $u$, it dominates the first type of anomaly, near  the sun, when
$\d\m\propto x^{-\a}$ at high $x$, for
$\a>3/2$.}$^{,}$\footnote{Note that this field differs from the
field of a central quadrupole, such as an oblate sun, which has the
same angular dependence but a different radial one.}
\par
Near the sun, the anomaly, as defined in eq.(\ref{diffa}), is a very
small difference between the much larger $\gf\_{in}$ and $\gf\^N$.
We have $|\gf_a|/|\gf\^N|=|q|(u/R\_M)[\az/(MG/u^2)]$; the last two
factors are very small (their product is $\sim 10^{-12}$ at $1\au$
from the sun), and we shall see that $q\sim 0.1$. Deducing the
anomaly directly from eq.(\ref{diffa}) by solving numerically for
$\f$ (e.g., using a non-linear Poison solver) could be very
challenging. Now that we identified the position dependence of the
anomaly, one can perhaps deduce $q$ by evaluating the anomaly
numerically at a larger distance from the Sun (e.g., at $10^3\au$
the relative strength of the anomaly is $\sim 10^{-4}$).
\par
Even though the effect is very small in the inner solar system, the
problem cannot be attacked by expanding the governing equations in
a small perturbation: There is always a region where the MOND
field differs greatly from the Newtonian one (see below), and the
ellipticity of the modified Poisson equation ties the different
regions inseparably.
\par
There is a useful way to envisage the anomaly based on our intuition
from Newtonian gravity, using the concept of ``phantom mass''
introduced in Milgrom (1986b).  This is the density
distribution, $\rp$, whose Newtonian field is the MOND anomaly. It is the mass density that a
Newtonist would attribute to dark matter, and is given by
 \beq \rp\equiv{1\over 4\pi G}\D\f_a={1\over 4\pi G}\div(\gf\_{in}-\gf\^N)={1\over 4\pi G}\div\gf\_{in}-\r.
\eeqno{phantom} We can also write \beq \rp={1\over 4\pi
G}\div\gf-\r={1\over 4\pi G}\div[(1-\m)\gf]. \eeqno{phantomi}
 The
exact MOND field in the system, $-\gf$, equals the sum of the
Newtonian fields of $\r$ and $\rp$, and of the constant (MOND)
external field.\footnote{The MOND external field can also be viewed
as the sum of the Newtonian galactic field, plus that of the
galactic phantom density.}
\par
In this way, we have identified the MOND anomaly as the full field
of a certain mass distribution, instead of as a very small
difference between two strong fields. This can be useful for deducing
the anomaly numerically, and for devising approximation schemes
that will bypass the inherent difficulty of deducing a small
difference of large quantities.

\par
Using the field equation (\ref{poisson}) we can write
 \beq \rp=-{1\over 4\pi
G\az}(\m'/\m)\grad|\gf|\cdot\gf+(\m^{-1}-1)\r, \eeqno{phantomu} or
in another useful form
 \beq \rp=-{\az\over 4\pi G}\ve\cdot\grad\U+(\m^{-1}-1)\r,
\eeqno{phantoma} where $\U(x)=\int L(x)dx$, with $L=x\m'/\m$  the
logarithmic derivative of $\m$,  $\ve$ is a unit vector in the
direction of $\gf$, and $x=|\gf|/\az$.
\par
We see from expression (\ref{phantomu}) that $\rp$ resides where
$\m$ differs from 1, namely, in low acceleration regions. There must
exist such a region, even when the external field itself is high:
Near the Sun the acceleration is pointing towards the sun. As we
move away in the direction of the galactic center, the acceleration
must eventually vanish at some point, and change sign to point in
the opposite direction. There is thus always a region $\P$ around
the critical point that is in the sub $\az$ regime, in which MOND
acts in full force, and in which $\rp$ is not negligible. This
region excludes a large volume around the Sun with our choice of
$\m$. Even with the limiting form of $\m$, where $\m=1$ for $x>1$,
the region $\P$, the distribution of $\rp$ in it, and thus the
anomaly that is felt near the sun, remain finite.
\par
The anomaly can be written as
 \beq \vg^{\anom}(\vR)=G\int{\rp(\vr)(\vr-\vR)\over|\vr-\vR|^3}\drt.
 \eeqno{nonono}
We thus identify\footnote{This can be used only when $\rp$ vanishes
at the origin, as I assume all along. Otherwise the $u$ integration
diverges at the origin and the integral is undefined. The quantity
this integral describes can still be finite when $\rp$ doesn't
vanish at the origin.}
 \beq q^{\anom}_{zz}=-{\left[dg^{\anom}_z\over dZ\right]_{\odot}}=
 G\int{d^3u\over u^3}\rp(\vu)(1-3\cos^2\t),
  \eeqno{nononum}
where $\vu=\vr-\vR_0$ is the position from the sun, and
$\cos\t=n_z\equiv u_z/u$.
\par
In Appendix A, I derive several useful, exact properties of the
phantom mass. First, I show that around the critical point there are
always regions of both positive and negative phantom density
(\ref{A3}) (this is an example of the general trend discussed in
Milgrom 1986b). Second, in \ref{A4} I derive the total phantom mass
$M_p$; it depends on $\eta$ and the form of $\m$.  (For the limiting
form of $\m$ we have $M_p=0$ identically; so the negative and
positive densities exactly cancel) . Third, I derive the column
density along the symmetry axis (\ref{A2}). It is independent of
$\eta$, and for the limiting form of $\m$ it is $\az/2\pi G$. Also,
I find [in \ref{A6}] that $\rp$ decreases asymptotically as
$u^{-3}$ (taking both signs at all radii); so this provides good convergence in eq.(\ref{nononum}).
\par
I also derive in the appendix [eq.(\ref{neratabar})] an expression
for the anomaly as an integral over $\gf$ (not involving derivatives
of $\gf$ as appear in $\rp$):
 \beq \vg^{\anom}(\vR)={1\over 4\pi}\int
 {(\m-1)\over |\vr-\vR|^3}[\gf-3(\gf\cdot\vn)\vn]\drt
 -{1\over 3}(1-\m_g)\vg_g,
\eeqno{neratteq} where $\m_g=\m(\eta)$ is the galactic value of
$\m$, and $\vn$ is a unit vector in the direction of $\vr-\vR$.
Although eq.(\ref{neratteq}) is equivalent to eq.(\ref{diffa}), it
is a better starting point for approximations. The great sensitivity
of the anomalous acceleration, as deduced from eq.(\ref{diffa}), to
errors in $\gf$ (or equivalently in $\gf\_{in}$), still remain in
expression (\ref{neratteq}), but here it is controllable. Errors in
the integration,  will cause $\vg^{\anom}(\vR)$ to vanish not at the
position of the sun, but at a somewhat different point off by
$\d\vR$. Even if $|\d\vR|/R\_M\ll 1$, the resulting errors in
$\vg^{\anom}(\vR)$ near the Sun could be very large. However, we can
eliminate this problem by considering directly\footnote{This
expression converges well near the Sun despite the $|\vu|^4$ factor,
and the divergence of $\gf$, because $\m-1$ vanishes quickly there
(and because the field becomes spherical, for which case the angular
integral vanishes). When $\d\m(x)$ vanishes faster than $x^{-3/2}$
for $x\rar\infty$, the $u$ integration is convergent near $u=0$, and
we can use this expression without further care. For $\d\m(x)\rar
x^{-3/2}$, the $u$ integral diverges logarithmically; however, the
angular integral vanishes, and the $z$ derivative of the force is
still finite at the sun. In this case, the anomalous field near the
sun is not the pure quadrupole, as I mentioned above. At infinity,
the convergence is also good, because there $\gf\rar -\vg_g+\grad\d$
with $\grad\d$ decreasing like $u^{-2}$. For the constant part the
angular integral vanishes, so there is convergence at least as
$d^3u/u^6$. See also eq.(\ref{nononum})}
 \beq q^{\anom}_{zz}=-{\left[dg^{\anom}_z\over dZ\right]_{\odot}}=
 -{1\over 4\pi}\int{(\m-1)\over |\vu|^4}
 [6\f_{,z}n_z+3\gf\cdot\vn(1-5n^2_z)]d^3u,
  \eeqno{anoqum}
where $\vn$ is now a unit vector in the direction of $\vu$. If in
eq.(\ref{anoqum}) we take $\vu$ in units of $R\_M$ and $\gf$ in units
of $\az$, the expression yields $q$.
\par
The phantom mass is only a visualization aid that I find very
convenient. We could have derived eqs.(\ref{neratteq}) and
(\ref{anoqum}), simply by starting from the identity (for functions
that decay fast enough at infinity):
 \beq \f_a(\vR)=-{1\over 4\pi}\int {\Delta\f_a(\vr)\over |\vr-\vR|}\drt=-{1\over 4\pi}\int {\div[(1-\m)\gf](\vr)\over |\vr-\vR|}\drt,  \eeqno{gumba}
(where the second equality comes from the definition of $\f_a$), and
proceed from there, integrating by parts, without mentioning phantom
matter. The concept is, however, also useful because my main
approximation is to replace the true phantom matter by a surrogate
one, which I do in section \ref{surrogate}.
\par
 To recapitulate,
eqs.(\ref{nonono})-(\ref{anoqum}) are only identities, useful here
in two ways: a. for calculating the quadrupole $q^{\anom}_{zz}$ as
an integral over ``large'' quantities, instead of as a very small
difference between much larger quantities, b. for devising
approximations for $q^{\anom}_{zz}$.

\par
To use eq.(\ref{anoqum}) to calculate the anomaly we need to solve
the nonlinear Poisson equation for the required choice of
interpolating function and an $\eta$ value, and use the resulting
field in eq.(\ref{anoqum}) (instead of subtracting the Newtonian
acceleration from the numerical solution very near the sun). It is
useful, however, to develop an approximation scheme that will permit
a quick calculation of $q$ for mass-production studies, since $\m$
and the relevant $\eta$ for the solar system are not exactly known,
and also for application to other precision systems, perhaps
involving more massive bodies. I shall then resort to an
approximation that will replace $\rp$ with a reasonable surrogate,
$\trp$, that is known in closed form. The surrogate anomaly can then
be written as an integral of a known function.
\section{Multi-potential theories
\label{multi}} It is interesting to study the EFE in the SS in other
modified-gravity formulations of MOND, such as multi-potential,
modified-Poisson theories. In these, the gravitational potential is
the sum of several potentials, each satisfying a modified Poisson
equation of its own. For example, the nonrelativistic limit of TeVeS
is a double-potential theory: the potential is the sum of a
Poissonian potential and one that is a solution of the nonlinear
Poisson equation, both having the true density as source.
Generalizing this, consider a nonrelativistic theory in which a
particle's equation of motion is $\va=-\grad\Phi$, with $\Phi=\sum_i
\f_i$, and with $\f_i$ determined from
 \beq \div[\m_i(|\gf_i|/\a_i\az)\gf_i]=4\pi\b_i G\r.  \eeqno{poissoni}
For concreteness sake, assume that in the limit $x\rar \infty$ all
$\m_i(x)$ go to constants\footnote{In principle we can have more
general behaviors, but I do not wish to delve into a general
discussion here.}, and take $\b_i$ such that all these constants are
1. (See relevant comments in Zhao \& Famaey 2006, Famaey \& al.
2007, Bruneton \& esposito-Farese 2007, and Milgrom \& Sanders 2008,
who deal with constraints on the form of the interpolating function
in TeVeS.) Also take $G$ to be the usual gravitational constant,
which means that $\sum_i\b_i=1$. The dimensionless coefficients
$\a_i$ can be absorbed into the definition of $\m_i$, but it is
convenient to keep them so that we can require a uniform low-$x$
behavior $\m_i(x)\rar x$. In this specific case, brought as an
example, we have in the very-low-acceleration limit
$|\gf_i|\ll\a_i\az$, for a spherical system, $|\gf_i|\approx
(\a_i\b_i)^{1/2}(\az|\gf\^{N}|)^{1/2}$, and $\az$ then plays its
usual role in phenomenology if we normalize it so that
$\sum_i(\a_i\b_i)^{1/2}=1$ (I assume that  $\a_i,\b_i\ge 0$, so all
potentials describe attractive gravity). If one of the $\a_i\rar 0$,
the corresponding $\m_i(x)\rar 1$ for all $x$; so, $\f_i$ satisfies
the linear Poisson equation $\div\gf_i=4\pi\b_i G\r$. For example,
the nonrelativistic limit of TeVeS would correspond to $\a_1\rar 0$,
and $\m_2=\m$ and $\b_2=\b$ still unspecified ($\b_1=1-\b$,
$\a_2=\b^{-1}$).
\par
Each of the sub-potentials $\f_i$ is solved for separately, each
generates its own cloud of ``phantom matter'' (which do not interact
with other clouds), each creating its own sub-anomaly
 \beq \vg^{\anom}_i=-\grad(\f_i-\b_i\f\^{N})  \eeqno{subanom}
($\f\^{N}$ is the standard Newtonian potential, calculated with the
standard $G$).
\par
To apply our results to the multi-potential case, we first have to
decompose the galactic field at the system's (e.g., the sun's)
position to its different components (for example, by solving the
full problem for the Galaxy's mass distribution, determining the
values of the sub-fields at that position): $\vg_g=\sum_i \vg_g^i$.
Then, remembering that the effective constants for each potential
are $G_i=\b_i G$, $\az^i=\a_i\az$, and
$R\_M^i=(\b_i/\a_i)^{1/2}R\_M$, we define
$\eta_i=|\vg_g^i|/\a_i\az$. We then calculate the dimensionless
quadrupole anomaly $q_i(\eta_i)$ using our results above and below.
We can also, as an approximation that is exact for a spherical
galaxy, use the algebraic relations
 \beq \vg_g^i=\b_i\n(\b_i |\vgN^g|/\a_i\az)\vgN^g,   \eeqno{miugas}
to determine $\vg_g^i$ from the Newtonian galactic field $\vgN^g$.
\par
The Sun is at a position in the Galaxy with a symmetric perspective.
Under these simplifying circumstances, all the $\vg_g^i$ must point
approximately in the same direction: towards the galactic center.
The anomaly's symmetry axis is then common to all the sub anomalies,
and we simply have, again, a biaxial, quadrupole anomaly with
$-2q^{\anom}_{xx}=-2q^{\anom}_{yy}=q^{\anom}_{zz}=q\az/R\_M$, and
 \beq q=\sum_i q_i(\eta_i)\a_i^{3/2}\b_i^{-1/2}. \eeqno{sumanom}
\par
For a planetary system placed at an arbitrary position in the
Galaxy, the directions of $\vg_g^i$ may be different. We then
straightforwardly add up the different quadrupole sub anomalies. The
result is still a traceless quadrupole; but it is, in general,
triaxial, with all three principal axes, and elements, determined by
the $q_i(\eta_i)$ and the respective directions of $\vg_g^i$.
\par
For TeVeS, the Poissonian field, $\f_1$, does not create an anomaly.
The galactic field $\vg_g^2$ defines the symmetry axis, even for an
arbitrary location in the Galaxy. The anomaly is thus always a
biaxial quadrupole, with $q=\b^{-2}q_2(\eta_2)$.

\section{The surrogate phantom density and its properties}
\label{surrogate} As an approximation, I shall substitute for  $\rp$
a surrogate phantom density, $\trp$, derived from the divergence of
the MOND field, as approximated by the algebraic relation
eq.(\ref{algebraic}). Call this field $\vg^*$, then
 \beq \trp= -{1\over 4\pi G}\div\vg^*-\r=
 -{1\over 4\pi G}\div(\vg^*-\vgN),    \eeqno{hutaf}
 with $\vg^*=\n(|\vgN|/\az)\vgN$, and
 \beq \vgN= -\eN\az\ve_z-{\msun G\over |\vu|^3}\vu=
 -\az(\eN\ve_z+{\hat\vu\over |\hat\vu|^3}),\eeqno{mugter}
where $\hat\vu=\vu/R\_M$ (the galactic field is in the $-z$
direction), and $\eN$ is the Newtonian value of $\eta$. We thus have
$\rp$ in closed form.\footnote{Note that I do not use $\vg^*$ as the
MOND field; i.e., I do not substitute it for $\gf$ for use in
eq.(\ref{diffa}). I only use it to obtain an approximation for
$\rp$. The Newtonian field of $\trp$; i.e., the surrogate anomaly
$\tilde\vg^{\anom}$, which is derivable from a potential, is not
equal to $\vg^*-\vgN$, even though the two fields have the same
divergence everywhere; $\vg^*$ is not, generally, derivable from a
potential. The two differ by a curl field.}
\par
All the results based on the surrogate density depend on the
Newtonian galactic field at the sun's position, $\vg\^N_g$, and do
not require knowledge of the MOND value of this field, $\vg_g$,
while the results using the exact theory depend on the latter alone.
The relation between these two accelerations depends on the galactic
mass distribution, which we don't know exactly (we can measure
$\vg_g$, but need a mass model to determine $\vg\^N_g$). We do know
that because the Sun is very near the galactic mid-plane, symmetry
dictates that the two are very nearly parallel. I assume in all that
follows that they are, and that their values are related by
$\eN=\mu(\eta)\eta$, which is exact for a spherical galaxy. This is
a good approximation near the sun, and, in any event, has little
consequence for us here, since it is only used to relate the value
of $\eN$ to be used in the surrogate-density approximation to the
value of $\eta$ appropriate for the sun: the small uncertainty in
$\eN(\eta)$ is rater immaterial.
\par
The surrogate phantom density $\trp$ is somewhat different from
$\rp$. However, $\trp$ does have properties (demonstrated in
Appendix B) that are rather similar to those of $\rp$, on which we
can found the hope that this is a reasonable approximation for
calculating the anomaly: First, $\trp$ has a  total mass similar to
that of $\rp$ (for the limiting form of $\m$ they both vanish)
(\ref{B1}). Second, $\trp$ has a similar pattern of sign
distribution (\ref{B2}). Third, the column density of $\trp$ along
the symmetry axis is identical to that of $\rp$ (\ref{B4}). Fourth,
the Newtonian field of $\trp$, like that of $\rp$, vanishes at the
position of the sun, and is thus also  a traceless, quadrupole near
the Sun (\ref{B3}).
\par
In some regards, the surrogate density does not quite mimic $\rp$.
For example, as can be seen from appendices \ref{A6}, and \ref{B7},
the angular distributions of their asymptotic behavior (for
$u\gg\eta^{-1/2}R\_M$) are different.

\par
As shown in \ref{A5}, substituting $\trp$ for $\rp$ is tantamount to
replacing $\gf$ in eq.(\ref{neratteq}) by $-\vg^*$. This gives for
the surrogate anomaly:
 \beq \tilde\vg^{\anom}(\vR)={1\over 4\pi}\int{(\n-1)\over
|\vr-\vR|^3}[\vgN-3(\vgN\cdot\vn)\vn]\drt-{1\over
3}(\n_g-1)\vg\^N_g. \eeqno{nettar}

Here $\n_g=\n(\eN)$ is the galactic value of $\n$ [$(\n_g-1)\vg\^N_g=(1-\m_g)\vg_g$].
 From this follows the expression for the dimensionless, surrogate,
anomalous quadrupole:
 \beq  \tilde q(\eta)=
 -{1\over 4\pi}\int{(\n-1)\over |\vu|^4}
 [6g\^N_zn_z+3\vgN\cdot\vn(1-5n^2_z)]d^3u,
  \eeqno{anoor}
with $\vgN$ from eq.(\ref{mugter}), now expressed in units of $\az$,
and $\vu$ in units of $R\_M$. We can write in spherical
coordinates\footnote{This expression can be used when $\d\m(x)$
vanishes faster than $x^{-3/2}$ for $x\rar\infty$ (equivalent to
$\n-1$ vanishing faster than $y^{-3/2}$). Otherwise, the $u$
integral diverges and we are not justified in choosing this order of
integration. For example, for the isolated, spherical case
($\eN=0$), this expression vanishes for any choice of $\m$, which is
a wrong result when $\d\m(x)$ vanishes as $x^{-3/2}$, or slower.}
 $$ \tilde q(\eta)={3\over 2}\int_0^{\infty}{du\over u^2}\int_{-1}^1 d\xi (\n-1)[\eN(3\xi-5\xi^3)+u^{-2}(1-3\xi^2)]=$$
 \beq = -3\int_0^{\infty}{du\over u^2}\int_{-1}^1 d\xi (\n-1)[\eN P_3(\xi)+u^{-2}P_2(\xi)], \eeqno{sphera}
where $P_{\ell}$ is the $\ell$s Legendre polynomial, and with
$\nu=\nu[(\eN^2+u^{-4}+2\eN u^{-2}\xi)^{1/2}]$  ($\xi=\cos\t$).
Because the angular integral vanishes when the factor $\n-1$ in the
integrand is $\xi$ independent, we can replace the 1 in $\n-1$ with
a function of $u$ alone, without changing the result. For example,
we can replace this 1 with the value of $\n$ for $\xi=0$, i.e., with
$\nu[(\eN^2+u^{-4})^{1/2}]$, or by the value of $\n$ at $\xi=\pm 1$:
$\nu(|\eN\pm u^{-2}|)$. With these, convergence is improved also at
large $u$, where $\n\rar \n(\eN)\not = 1$. Changing variables to
$v=u^{-1}$ we can also write
  \beq  \tilde q(\eta)= -3\int_0^{\infty}dv\int_{-1}^1 d\xi (\n-1)[\eN
P_3(\xi)+v^2P_2(\xi)], \eeqno{spheravv}
 with $\nu=\nu[(\eN^2+v^4+2\eN v^2\xi)^{1/2}]$.
\par
In Appendix \ref{C}, I consider further one particular case: the
limiting form $\m(x)=x$ for $x\le 1$, and $\m(x)=1$ for $x>1$. This
is useful pedagogically, since it lands itself to further
approximation that will lead to a closed expression for the
surrogate anomaly. This can be used, for example, as a touchstone
for checking numerical results. It also affords a simple description
of the phantom matter; and, it also seems to give a lower limit to
the anomaly for a large class of interpolating functions.
\par
For this limiting form of $\m$, with $\eN>1$ (which is relevant for
the SS), we have a well demarcated region $\tP$ outside which
$\trp=0$: We have at large distances in all directions $|\gf|\rar
|\vg_g|>\az$, which takes us outside $\tP$; so $\tP$ is bounded in
all directions. The boundaries of $\tP$ can then be obtained in
closed form, and are described in \ref{B5}. For $\eN<1$, or for
choices of $\m(x)$ that reach the value 1 only asymptotically at
high $x$, the region $\tP$ is not bounded.
\par
The further approximation that I apply to the limiting case can be
justified in the limit of large $\eta=\eN$, and leads to [see
eq.(\ref{anomasa})]
 \beq \hat q(\eta)=-\eta^{-7/2}/30.
 \eeqno{anoma}

\section{Results for the dimensionless quadrupole anomaly}
\label{results} I use eq.(\ref{sphera}), or eq.(\ref{spheravv}), to
compute the dimensionless, surrogate quadrupole anomaly $\tilde
q(\eta)$ for various forms of $\m(x)$ and different $\eta$
values.\footnote{As a check of the numerics I also use
eq.(\ref{nettar}) to calculate the $z$ component of the anomalous
acceleration along the $z$ axis, near the sun, to ascertain that is
vanishes at the sun's position, and to check how good the quadrupole
approximation is away from the sun. Both aspects check well; for
example, the departure from a pure quadrupole behavior is, as
expected, of the order or $u/R\_M$, only about one percent within
$80 \au$.} I also wrote a simple nonlinear-Poisson solver (based on
the detailed blueprints described in Milgrom 1986a) to calculate $q$
for several control cases from the exact expression (\ref{anoqum}).
\par
 I consider three one-parameter families of interpolating
functions, defined and discussed in Milgrom \& Sanders (2008), and
checked to give good results in rotation curve analysis in the
transition region. They also have a rapid decline of $\dmx$ at high
$x$, with a negligible anomaly resulting from the finiteness of
$\dmx$. The first family is
 \beq \mu_\a(x)={x\over (1+x^\a)^{1/\a}}, \eeqno{v}
with the corresponding
 \beq \nu_\a(y)=\left[{1+(1+4y^{-\a})^{1/2}\over 2}
 \right]^{1/\a},
 \eeqno{va}
 which I use with $\a>3/2$.
The limiting form of $\m(x)$ corresponds to very large $\a$.
\par
The second and third families are defined by their $\nu(y)$
functions:
 \beq \tnu_{\a}(y)=(1-e^{-y})^{-1/2}+\a e^{-y}, \eeqno{vi}
and
 \beq
 \bnu_{\a}(y)\equiv(1-e^{-y^\a})^{(-1/2\a)}+(1-1/2\a)e^{-y^\a},\eeqno{vii}
 with the corresponding $\tmu_{\a}(x)$, and $\bmu_{\a}$. The choice of $\a$ permits a very
fast transition to the Newtonian regime near $y=1$. These choices
are  only indicative; they are neither fully representative, nor exhaustive.
\par
Table \ref{table1} gives the numerical results for $-\tilde q$. For
some entries I also give in parentheses the value obtained from the
numerical solution of the modified Poisson equation via
eq.(\ref{anoqum}).
\par
The value of $\eta$ for the galactic field in the solar neighborhood
is given by
 \beq \eta\approx 1.9\left({V\_{\odot}\over 220\kms}\right)^2
 \left({R\_{\odot}\over 8\kpc}\right)^{-1}
 \left({\az\over  10^{-8}\cmss}\right)^{-1}, \eeqno{eta}
  with $V\_{\odot}$ the orbital velocity of the Sun in the
Galaxy, and $R\_{\odot}$ its orbital radius.\footnote{The value of
$\az$ is determined from several of its appearances in the
phenomenology to be around $10^{-8}\cmss$. For example, Begeman,
Broeils, \& Sanders (1991) found from rotation curve analysis
$\az\approx 1.2\cdot 10^{-8}\cmss$. Recently, Stark, McGaugh, \&
Swaters (2009) found that gas dominated galaxies satisfy the
mass-asymptotic-rotational-velocity relation predicted by MOND--aka
the baryonic Tully-Fisher relation--($M=\az^{-1}G^{-1}V^4_{\infty}$)
with $\az=1.2\cdot 10^{-8}\cmss $, and a range of $(0.7-2)\cdot
10^{-8}\cmss$.}$^,$\footnote{When dealing with multi-field theories
rather different $\eta$ values may appear.} The relevant range for
our problem is thus roughly $1\lesssim \eta\lesssim 2$. In the table
this gives a range $0.01\lesssim -\tilde q\lesssim 0.3$. The
interpolating functions in the last two rows are rather less likely;
if we omit those the range shrinks to $0.07\lesssim -\tilde
q\lesssim 0.3$.
\par
We see that the surrogate approximation works rather well, and by and large produces $\tilde q$ values
that do not differ  from $q$ by much. We also see that that expression (\ref{anoma})
is a rather good approximation, in the limiting case,
down to $\eta$ values of 1.5, even though it can be justified as an
 approximation only for large values of $\eta$.
\par
Notice that $|\tilde q|$ decreases for increasing  $\eta$, when
$\eta \gg 1$, because the phantom matter becomes less and less
important; $|\tilde q|$ also decreases with decreasing $\eta$, for
$\eta< 1$, because the phantom matter becomes more and more
spherical. The maximum effect seems, in fact, to occur for the range
of $\eta$ values around that for the galactic field near the sun.
\begin{table}

\begin{center}

\begin{tabular}{|c  |c  |c  |c  |c  |c  |c  |c  |}
  \hline
  $\eta\rar$    & $0.5$ &$1$&$1.5$&$2$   &$3$&$5$&$10$ \\
  \hline
  $\bmu_2$& $ 3.4\cdot 10^{-2}  $ &$0.12$& $ 0.20 $&$ 9.8\cdot 10^{-2} $ &$2.5\cdot 10^{-2}$& $ 3.8\cdot 10^{-3} $&$ 3.1\cdot 10^{-4} $     \\
  $\bmu_{1.5}$   & $ 3.5\cdot 10^{-2}  $ &$ 0.11 $& $0.19$ &$ $0.16$ $&$ 6.3\cdot 10^{-2} $&$ 1.1\cdot 10^{-2} $ &$ 9.2\cdot 10^{-4} $  \\
  $\tmu_{.5}$ & $ 3.6\cdot 10^{-2}  $& $0.11$ & $0.17$ &$ 0.21 $ &$ 0.18 $&$ 9.0\cdot 10^{-2} $ &$ 1.2\cdot 10^{-2} $ \\
     & & ($0.16 $)&($0.21 $) &($0.24 $) &($0.20 $) &&   \\
  $\tmu_{1}$ & $ 3.2\cdot 10^{-2}  $ &$ 0.11 $ &$ 0.21 $&$ 0.28 $ &$ 0.28$&$ 0.13 $ &$1.8\cdot 10^{-2}$ \\
  $\m_2$ & $ 3.9\cdot 10^{-2} $&$ 8.8\cdot 10^{-2}  $ &$0.11 $ &$ 0.12 $ & $0.12$&$ 0.11 $ & $8.2\cdot 10^{-2}$  \\
      & &($ 0.10 $) &($0.12 $) &($ 0.12$) & ($0.12$)&&   \\
  $\m_3$ & $ 4.0\cdot 10^{-2} $ &$8.1\cdot 10^{-2}$& $7.9\cdot 10^{-2}$&$6.5\cdot 10^{-2}$&$4.4\cdot 10^{-2}$&$ 2.3\cdot 10^{-2} $ &$ 8.8\cdot 10^{-3} $  \\
  $\m_{\infty}$& $ 4.3\cdot 10^{-2} $&$ 4.4\cdot 10^{-2} $&$ 8.8\cdot 10^{-3} $&$ 3.1\cdot 10^{-3} $ &$ 7.3\cdot 10^{-4} $&$ 1.2\cdot 10^{-4} $&$1.1\cdot 10^{-5}$  \\

& &$
(5.3\cdot 10^{-2}) $&$ (1.1\cdot 10^{-2}) $&$ (4.6\cdot 10^{-3}) $ &   &    & \\
  $eq.(\ref{anoma})$ & $ 4.3 $&$3.3\cdot10^{-2} $&$ 8.1\cdot10^{-3} $&$ 2.9\cdot10^{-3} $ &$7.1\cdot10^{-4}$&$ 1.2\cdot10^{-4} $    &$1.1\cdot10^{-5}$   \\
   \hline
\end{tabular}
\caption{Values of the dimensionless, surrogate,  quadrupole anomaly
$-\tilde q$ for various choices of the interpolating function and
various values of $\eta$. In parentheses are shown some values of
$-q$ calculated using numerical solutions of the exact modified
Poisson equation. \label{table1}}

\end{center}
\end{table}
\section{The effect of the anomalous quadrupole on planetary motions}
\label{solar} For $|\vu|\ll R\_M$, the quadrupole anomaly due to the
MOND EFE is of the same form as that produced by tidal effects of a
static distant mass\footnote{At radii of order $R\_M$ or larger the
effect is very different from tidal effects, being the Newtonian
field of $\rp$. So, for example, it decreases beyond $\sim R\_M$.}
(such as a molecular clouds, a nearby star, etc.). For example, a
solar mass star at a distance  $d$ is expected to give
$q^*_{zz}\approx -4\cdot 10^{-5}(d/\pc)^{-3}\az/R\_M\approx -4\cdot
10^{-4}(q/0.1)^{-1}(d/\pc)^{-3}q^{\anom}_{zz}$. The tidal effect of
the Galaxy  is dominated by the gradient in the direction
perpendicular to the disc, call it the $x$ axis, and we have for its
quadrupole component $q\^{tide}_{xx}\approx 4\pi \r_0$, where
$\r_0\approx 0.076 \msun/\pc^3$ (Cr\'ez\'e \& al. 1997) is the
midplane dynamical density of the Galaxy at the sun's position. We
thus have $q\^{tide}_{xx}\approx 3\cdot 10^{-4}(q/0.1)^{-1}
q^{\anom}_{zz}$. The other two quadrupole components of the galactic
tidal field are about an order of magnitude smaller. Such effects
are thus much smaller than the EFE quadrupole anomaly.
\par
One may wonder what the MOND effects are of the planets on each
other. Beside the standard physics, mutual effects, which are always
taken into account, there are new MOND effects here as well. Again,
some of these have to do with the remaining small value of $1-\m(x)$
at high $x$, which we ignore here. The effects that remain even with
the limiting form of $\m$ can, again, be described as the Newtonian
effects of the phantom matter.  For any given configuration of the
SS there are several critical points, each begetting its own cloud
of phantom matter; for example, one near each planet due to its
competition with the Sun (Bekenstein \& Magueijo 2006 consider the
potential for probing such regions directly). There is also some
distortion of the main phantom matter distribution near the
solar-galactic critical point. I intend to discuss the problem of a
few-body system elsewhere. Here I only mention that these effects
are negligible in the present case.

\par
To test for the presence of the quadrupole anomaly, and constrain
its parameters, one needs to incorporate the anomalous force in the
equations of motion in programs that fit for the observed motions of
the planets. Today, the direction to the galactic center, which we
take to define our symmetry axis, is only about $6$ degrees away
from the ecliptic. To first approximation, it can thus be assumed
that our symmetry axis is in the planetary orbital plane, and take
the $x$ axis to also lie in this plane.\footnote{This is not a good
assumption for Pluto and Icarus, which I also consider. It also
breaks down at other phases in the galactic rotation.} Then, the
anomalous acceleration in the orbital, $x-z$ plane, is
 \beq\vg^{\anom}={q^{\anom}_{zz}\over 2}(x,-2z),  \eeqno{plana}
at position $\vu=(x,z)$ from the sun.
\par
To get an idea of the size of the expected effects of the anomaly on
the planetary motions, consider, for example, the anomalous
perihelion advance rate it produces. Let
  \beq u={a(1-e^2)\over 1+e~ \cos\psi},  \eeqno{orba}
be the unperturbed Keplerian orbit of a planet, with the azimuthal
angle $\psi$ (which increases with time) measured from perihelion;
($a$ and $e$ are the semi-major axis and eccentricity,
respectively).
\par
One can use the Gauss equations to calculate the variation of the
orbital elements under the action of a given acceleration field
(Brouwer \& Clemence 1961). For this, one needs the instantaneous
radial, tangential, and normal components of the acceleration,
$g^{\anom}_r,~g^{\anom}_t,$ (in the direction of increasing $\psi$),
and $g^{\anom}_n$, respectively. In our case they are
 \beq g^{\anom}_r={q\az u\over 2R\_M}[1-3\cos^2(\psi-\psi_0)], \eeqno{rata}
 \beq  g^{\anom}_t={3q\az u\over 2R\_M}\sin(\psi-\psi_0)\cos(\psi-\psi_0), \eeqno{tata}
and I take $g^{\anom}_n$ to vanish (in the ecliptic it is about an
order smaller than the other components). Here $\psi_0$ is the value
of $\psi$ on the positive $z$ axis.
 With $g^{\anom}_n=0$, there is no precession of the
orbital plane, and the instantaneous perihelion advance, for
example, can be calculated as
 \beq \dot\o={(1-e^2)^{1/2}\over nea}[-g^{\anom}_r\cos\psi
 +g^{\anom}_t(1+{u\over p})\sin\psi], \eeqno{omega}
where $\o$ is the longitude of perihelion of the perturbed
trajectory, $n=(\msun G/a^3)^{1/2}$, and $p=a(1-e^2)$. To get the
mean perihelion advance rate we average this rate over a period. I
find
 \beq \langle\dot\o\rangle=-{q\az\over nR\_M}
 \k(e)\l(\psi_0),  \eeqno{perihelion}
 where
 \beq \k(e)=(1-e^2)^3\left({1+s^2\over 1-s^2}\right)^5,~~~~~
 \l(\psi_0)={3\over 8}[1+5\cos(2\psi_0)],  \eeqno{lambda}
with $s\equiv[1-(1-e^2)^{1/2}]/e$.

So,
 $$\langle\dot\o\rangle\approx 4.4\cdot 10^{-20}\k(e)\l(\psi_0)\left({-q\over
0.1}\right)\left({\az\over 10^{-8}\cmss}\right)^{3/2}\left({a\over
\au}\right)^{3/2}
 {\rm s}^{-1}=$$
 \beq =2.8\cdot 10^{-5}\k(e)\l(\psi_0)\left({-q\over
0.1}\right)\left({\az\over 10^{-8}\cmss}\right)^{3/2}\left({a\over
\au}\right)^{3/2}
 \aspc.
  \eeqno{perihel}
Note that, depending on $\psi_0$, the effect can take up both signs,
as $\l(\psi_0)$ can vary between $-3/2$ and $9/4$.
\par
 I also calculated the predicted rate of secular changes in the semi-major axis due to the
 quadrupole anomaly and found that it vanishes for all values of $e$
 and $\psi_0$.
\par
The $\k\l$ values, and the predicted anomalous perihelion precession
rates for the planets, and for Icarus, are shown in Table
\ref{table2}.
\par
Also shown, under $|\Delta\dot\o|$, are some indicative intervals
for $\dot\o$  from published best-fit results for the precession
rates, obtained through fitting to specific conventional models.
They are shown as an indication of the level of accuracy that has
been reached for the different planets.  The tightest values are for
Earth and Mars and, for our reference value of $q=0.1$, are $\sim
8-10$ times larger than the expected rates for the two planets. So,
parameter values that give $q\sim 0.3$, and $\az=1.2\cdot
10^{-8}\cmss$, could already be only a factor 2-3 below these
intervals.\footnote{Fienga \& al. (2009) quote Pitieva as giving a
value of $(-6\pm 2)\cdot 10^{-3}\aspc$, for the residual perihelion
precession of Saturn, which might be viewed, formally, as a positive
detections of an anomaly. Other analyses they quote give $(-10\pm
8)\cdot 10^{-3}\aspc$. Such an anomaly could be explained, e.g., by
the harmonic anomaly I discuss below. However, since these authors
seem not to view these results as definitely indicating a detection
of an anomaly, I took it as an indicative accuracy of
$|\Delta\dot\o|\sim 10^{-2}\aspc$ for Saturn.}
\par
For the outer planets, the effect is expected to be larger, but so
are the measurement errors, at present. So the prospects of first
constraining the anomalous MOND EFE would seem to come if the
measurements for the outer planets can be improved.

\par
It is interesting to compare the above results with what we get from
the harmonic anomaly predicted in the inner SS--even without an
external field--if the asymptotic behavior of the interpolating
function is $\m(x)\rar 1-Ax^{-3/2}$. In this case we get generically
in MOND, {\it an additional} anomaly near the sun, with
  \beq g^{h}_r={-A\az u\over R\_M}, \eeqno{ratats}
and with $g^{h}_t=0$. It is thus of the same order, and has the same
distance dependence as the quadrupole anomaly. Using these in the
Gauss equations we get a secular perihelion rate:
  \beq \langle\dot\o^h\rangle=-{3A\az\over 2nR\_M}\k(e),  \eeqno{perihon}

or,
 $$\langle\dot\o^h\rangle\approx -6.5\cdot 10^{-19}A\k(e)\left({\az\over 10^{-8}\cmss}\right)^{3/2}\left({a\over
\au}\right)^{3/2}
 {\rm s}^{-1}=$$
 \beq =-4.1\cdot 10^{-4}A\k(e)\left({\az\over 10^{-8}\cmss}\right)^{3/2}\left({a\over
\au}\right)^{3/2}
 \aspc.
  \eeqno{perel}
I give in Table \ref{table2} the predicted rates for this effect for
the value $A=2/3$, which corresponds, for example, to the form
$\m_{3/2}=x(1+x^{3/2})^{-2/3}$, which has the appropriate asymptotic
form.

\begin{table}

\begin{center}

\begin{tabular}{|c  |c  |c  |c  |c  |c  |c  |c  |c |c |c|}
  \hline
  $    $    & Merc. &Ven.&Ea.&Mar. &Jup.&Sat.&Uran.&Nept.&Plut.& Icar. \\
  \hline
  $\k\l$& $ 2.1 $ &$.38$& $ 1.98 $&$ -1.0 $ &$-1.1$& $ 2.2 $&$ -1.5 $ &.56 &  .50 &  $ -.18$\\
  $\langle\dot\o\rangle$ & $ .13 $ &$ .065$& $.53$ & $-.53$ &$ -3.8 $&$ 18 $ &$ -34 $ & $26$& $31$& $ -.054$ \\
  $|g^{\anom}|$ & $ .048 $& $.094$ & $.13$ &$ .20 $ &$ .67 $&$ 1.2 $ &$ 2.5 $ &$3.9$ & $4.8$ & $.14 $\\
  $-\langle\dot\o^h\rangle$ & $ .60 $ &$ 1.7$& $2.7$ & $5.1$ &$ 32 $&$ 80 $ &$ 230 $ & $450$& $590$ & $1.7$ \\
  $|g^h|$ & $ .32 $& $.62$ & $.86$ &$ 1.3 $ &$ 4.5 $&$ 8.2 $ &$ 17 $ &$26$ & $32$  & $.93 $\\
  $|\Delta\dot\o|$ & $ 50 \^{(a)} $& $5^{(b)}$ & $4\^{(a)}$ &$ 5\^{(a)} $ &$2000\^{(c)}  $&$ 100 \^{(c)}$ &$2\cdot10^5$ $ \^{(c)}$ &$2\cdot10^5$ $ \^{(c)}$ & $ $\\

    \hline
\end{tabular}

\caption{Characteristics relating to the perihelion precession rates
produced by the EFE quadrupole anomaly, and the harmonic anomaly:
$\k\l$ as defined in eq.(\ref{lambda}); the predicted precession
rate $\langle\dot\o\rangle$ for the quadrupole anomaly (calculated
assuming that the orbit is in the ecliptic--not a very good
approximation for Pluto and Icarus.); the mean anomalous quadrupole
acceleration on the orbit, $|g^{\anom}|=|q_{zz}|a$; the precession
rate due to the harmonic anomaly $\langle\dot\o^h\rangle$; its mean
acceleration $|g^h|=A\az a/R\_M$; and an indication of existing
tightness of conventional fits, $|\Delta\dot\o|$, that I was able to
find in the literature: from (a) Pitjeva (2005), (b) Pitieva as
quoted in Fienga \& al. (2009), (c) Fienga \& al. (2009). All
precession rates in units of $10^{-4}\aspc$. All accelerations in
units of $10^{-12}\cmss$. All values are calculated for
$\az=10^{-8}\cmss$, $q=-0.1$, and $A=2/3$. \label{table2}}

\end{center}
\end{table}

\section{Discussion}
\label{discussion}
\par
In the framework of the modified-Poisson formulation of MOND, one
expects an anomalous quadrupole acceleration
$g^{\anom}_i(\vu)=-q^{\anom}_{ij}u^j$, at position $\vu$ with
respect to the sun, with $q^{\anom}_{ij}$ diagonal, axisymmetric,
and traceless:
$-2q^{\anom}_{xx}=-2q^{\anom}_{yy}=q^{\anom}_{zz}=q(\eta)(\az/R\_M)$.
Interestingly, this anomaly remains finite even if the MOND
interpolating function is arbitrarily close to unity at high
accelerations, despite the fact that the accelerations near the sun
are $\gg\az$. I find numerically $|q|\sim 0.1$, and this corresponds
to an anomalous acceleration $\sim 10^{-5}\az$ for the inner
planets, but $\sim 10^{-4}\az$ for Saturn, and even larger for the
yet farther planets. These do not conflict with published SS bounds
on anomalous perihelion precession, but useful constraints seem to
be within reach.
\par
This effect can be conveniently envisaged as the tidal effect on the
solar system, of the imaginary phantom mass that, according to MOND,
hovers around the sun.
\par
In all the examples I studied, $q$ turned out to be negative
(repulsion from the Sun along the $z$ axis and attraction in the
perpendicular direction). I was not able to prove that this is
always the case from basic properties of $\m$. This question of the
sign of $q$ requires a more systematic, extensive study.
\par
Our results here also apply to the unusual sort of gravitationally
polarizable matter proposed by Blanchet \& Le Tiec (2008, 2009) as a
model for MOND, as it reproduces the potential of the nonlinear
Poisson equation.
\par
I find that the strength of the effect is sensitive to the form of
the interpolating function $\m(x)$ in the MOND-Newtonian transition
region ($x\sim1$). The constraints we have on this aspect of $\m$
comes mainly from rotation-curve analysis. Different studies--e.g.
by Famaey \& Binney (2005), Zhao \& Famaey (2006), Sanders \&
Noordermeer (2007), and Milgrom \& Sanders (2008)--have shown that
rotation-curve analysis can, at present, be used to constrain $\m$
in its transition region. However, all existing studies make use of
the prediction of modified-inertia formulations. These establish a
universal, algebraic relation of the form (\ref{algebraic}) between
the Newtonian and MOND accelerations {\it for circular orbits}
(Milgrom 1994). For consistency, such attempts to constrain $\m$ for
use in the present context should use the predictions of the
modified Poisson formulation (or a multi-potential theory, if we are
dealing with one). This is not an easy task since this theory does
not provide an easy-to-use formula, but would require for each
galaxy model, and for each choice of parameters, a separate
numerical solution of the nonlinear Poisson equation. Brada \&
Milgrom (1995) did show that the differences in prediction of the
two formulations are not very large, but they are of the same
magnitude as the differences produced by different interpolation
functions.
\par
Interesting constraints on the form of $\m$ in the transition region
will also come from studies of the dynamics perpendicular to the
galactic disk at the solar position (Milgrom 1983, and see a recent
study of the prospects of such analysis in Bienaym\'e \& al. 2009).
\par
We thus cannot, at present, predict the exact strength of the
effect, for lack of exact knowledge of the interpolating function
and the value of $\eta$. However, the presence of the effect can be
easily identified if we can observe the anomaly in the motion of more than one planet: The
effect for all planets hinges on only one parameter, $q$, and the
dependence of the effect on orbital parameters is very distinctive.
The anomalous acceleration, and the precession rate, increases with
orbital radius in a prescribed way. The precession rate depends
also on the orientation of the orbit relative to the galactic center
in a known manner [through $\l(\psi_0)$], which distinguishes it from a
spherical anomaly (or from an aligned, axisymmetric one). In contrast, a central
quadrupole, such as is produced by an oblate sun, for example, has
very different characteristics: its effect decreases sharply with increasing
orbital radius, and the perihelion precession rate it produces does
not depend on the orientation of the orbital major axis with respect
to the quadrupole axis (our $\psi_0$) when that axis is in the
orbital plane (or perpendicular to it).
\par
The MOND anomalous acceleration becomes of order $\az$ at radii
$\sim R\_M$ and beyond\footnote{At these radii, the field no more has the
quadrupole form that which obtains near the sun.}, and could have
important effects on object in the outer solar system, such as the
long period comets. MOND could introduce completely new insights
into the study of the formation, structure, and evolution of the
Oort cloud, and shed new light on possible mechanisms for directing
Oort cloud object into the inner solar system. But considerations of
these sort are beyond the scope of my present discussion.
\section*{Acknowledgements}

 This research was supported by a
center of excellence grant from the Israel Science Foundation. I
thank Luc Blanchet and Gilles Esposito-Farese for reading the
manuscript, and for useful information.
\appendix
\section{Properties of the phantom mass}
\label{A}
There are several properties of the phantom mass that we
can derive analytically.

\subsection{The Newtonian field of the phantom mass vanishes at the position of the sun}
\label{A1} This is a simple corollary of a more general result: If
$\r(\vr)$ is a confined density distribution of total mass $M$, in a
constant, background (MOND) field $\vg_g$, then the force acting on
this distribution is $M\vg_g$ (Bekenstein and Milgrom 1984). On the
other hand this force is also $M\vg_g$ plus the force due to the
phantom mass. It follows then that the total anomalous force on the
true mass $\r$ always vanishes exactly. This is a generalization to
the case of an external field of the following observation: A mass
distribution does not self propel in either Newtonian gravity or in
the modified-Poisson formulation. This means that the difference of the net MOND force and
Newtonian force on the totality of (true) mass must vanish. But this
difference is just the  Newtonian force of the phantom matter
on the true matter, which must then vanish. The phantom matter does,
of course, exert differential forces on different parts of the true
mass--these are the MOND corrections to the inter-body forces in the
system--but they all add up to zero.
 In our present configuration the force on
the Sun is $\msun\vg_p(0)$, where $\vg_p(0)$ is the Newtonian field
of the phantom mass at the position of the sun, and it must thus
vanish. In the very near vicinity of the Sun the field of the
phantom mass--i.e., the anomaly we are after--is predominantly a
quadrupole.
\subsection{The column density along the symmetry axis}
\label{A2}
 On dimensional grounds, the phantom column density along
the $z$ axis can be written as
 \beq \S_p(0)={\az\over 2\pi G}\c(\eta). \eeqno{sumsu}
Integrating expression (\ref{phantoma}) for $\rp$ along the symmetry
axis in three segments between which $\ve$ changes sign: from
$-\infty$ to the critical point, from there to the sun's position, and
from there to $\infty$, we get
 \beq  \c=\U(\infty)-\U(0)=\int_0^{\infty}L(x)dx
 \eeqno{surface}
[$L(0)=1,~L(\infty)=0$]. The dimensionless factor $\c$
is thus independent of $\eta$ and depends only on the choice of
$\m$.
  For the limiting form of
$\m(x)$ we have $\U(\infty)-\U(0)=1$, hence in this case the column
density is exactly $\az/2\pi G$. For $\m_2$ we have $\c=\pi/2$. Note
also that $\c$ diverges if $\dmx$ behaves at large $x$ as $x^{-1}$
or slower.

\subsection{The phantom density takes up both signs near the critical point}
\label{A3}
 At the critical point $|\gf|$ vanishes, so it must be
increasing in all directions emanating thence. Take any surface,
$\S$, of constant $|\gf|$ that surrounds the critical point, but
excludes the sun. Applying Gauss's theorem to the modified Poisson
eq.(\ref{poisson}) for the volume enclosed by $\S$, which is devoid
of (true) mass, gives $\int_{\S}\gf\cdot\vds=0$. Thus, since
$\grad|\gf|$ is parallel to $\vds$ everywhere on the surface,
$\grad|\gf|\cdot\gf$ must take up both signs on every such surface.
Expression (\ref{phantomu}) for $\rp$ then tells us that so must
$\rp$.

\subsection{The total mass}
\label{A4}
 The total phantom mass is $M_p=\msun\b(\eta)$. The
dimensionless factor, $\b$, can be found by applying Gauss theorem
to eq.(\ref{phantomi}) in a spherical volume whose surface $S$ tends
to infinity.
 \beq M_p={1\over 4\pi G}\int_S(1-\m)\gf\cdot\vds.  \eeqno{gauss}
Here $\gf$ is the MOND field; its asymptotic form is known
analytically and given in  eq.(32) of Bekenstein \& Milgrom (1984).
Using this expression one finds
 \beq \b(\eta) = \m_g^{-1}L_g^{-1/2}\sin^{-1}\left({L_g\over
 1+L_g}\right)^{1/2}-1,
\eeqno{asym} where $L_g=L(\eta)$, $\m_g=\m(\eta)$.
\par
With the limiting form of $\m$, and $\eta>1$, we have $\m_g=1$ and
$L_g=0$; so the total phantom mass vanishes (but not $\rp$, of
course). It is easy to see this directly. In this case $\P$ is
bounded, and excludes the sun. Outside its boundary $\m=1$. So, on
any boundary, $\S$ surrounding $\P$ we have $\m\gf=\gf$. The surface
integral of $\m\gf$ vanishes, if $\S$ excludes the sun, as it is
proportional to the true mass in $\P$, and this means that
$\int_{\S} \gf\cdot\vds=4\pi G M_p$ also vanishes.
\par
If $\m$ does not have the limiting form, $M_p$ is finite. For
example for $\m=\m_2=x/(1+x^2)^{1/2}$, we have for $\eta=2$, $\b
\approx 0.05$, and for $\eta=1$, $\b\approx 0.23$. The phantom mass
is always non-vanishing if $\m(\eta)<1$. But note that $M_p$ in
itself is not an indication for the strength of the anomalous
quadrupole near the sun.
Taking an extreme case, when $\eta\rar 0$, we have $M_p\rar\infty$
(reflecting the fact that it then constitutes an asymptotically
isothermal phantom halo around the central mass). But the quadrupole
anomaly actually goes to 0 in this case as the phantom mass forms a
spherical cavity around the sun.
\par
All the above applies to a single star in a constant field. In
reality, the phantom masses of a star cuts off as it runs into those
of other stars. It is easy to see that the phantom mass of a galaxy
is just the totality of phantom masses of individual constituents;
e.g., by applying Gauss theorem to a large volume written as the
union of its sub volumes. One can then mimic the effects of MOND
with a smoothly distributed dark matter, but only when the dynamics
is probed with very course graining such as in rotation curve
analysis. Unlike the hypothetical CDM, the MOND phantom matter is
highly granular on the scales of distances between object in the
galaxy, with regions of negative density, etc.. This, including the
EFE we discuss in this paper, are not mimicked by CDM.
\subsection{Space integrals weighted by $\rp$}
\label{A5} One encounters various quantities of the form
 \beq F=\int\rp(\vr)f(\vr)\drt,  \eeqno{gushma}
where $f(\vr)$ can be any tensorial function of $\vr$. Examples are
the various multipoles of the phantom mass, and its Newtonian
potential and acceleration fields. Here I
derive useful expressions for such quantities writing them as
integrals over $\gf$ itself, instead of $\rp$, which involves second
derivatives of $\f$. These are quite useful for numerical
evaluations.
\par
Using expression (\ref{phantomi}) for $\rp$ we can write
 \beq F= {1\over 4\pi G}\int f(\vr)\div[(1-\m)\gf]\drt.  \eeqno{bushla}
Integrating by parts and applying Gauss's theorem
 \beq F= {1\over 4\pi G}\int f(\vr)(1-\m)\gf\cdot\vds-
 {1\over 4\pi G}\int (1-\m)(\gf\cdot\grad)f(\vr)\drt.  \eeqno{bushlaba}
We have already had one example of such an expression where $F$ is
the total phantom mass, with $f=1$. The second integral then
vanishes, and the surface integral at infinity can be evaluated for
the case of a finite external field (for $\vg_g=0$ the surface
integral diverges). In general, we do not have a prescription for
calculating the surface integral. However, there are cases when we
can evaluate it. For example, if $\m$ takes its limiting form, and
if $\eta>1$, the surface integral vanishes identically over any
surface surrounding $\P$, and we can write
 \beq F= {1\over 4\pi G}\int_{\P} (\m-1)(\gf\cdot\grad)f(\vr)\drt.
  \eeqno{bkiushla}
\par
In this case we get for the dipole moment, with $f=\vr$
 \beq \vD={1\over 4\pi G}\int_{\P}(\m-1)\gf\drt. \eeqno{nerata}
 For the quadrupole we have from eq.(\ref{bkiushla}), with
$f=-(r^2\d_{ij}-3r_ir_j)/2$,
  \beq Q_{ij}=-{1\over 4\pi
G}\int_{\P}(\m-1)[\vr\cdot\gf\d_{ij}-{3\over
2}(\f_{,i}r_j+\f_{,j}r_i)]\drt, \eeqno{nerataba}
 where $\ve$ is a unit vector in the direction of $\gf$.
In particular \beq Q_{zz}=-{1\over 4\pi
G}\int_{\P}(\m-1)(\vr\cdot\gf -3z\f_{,z})\drt. \eeqno{neratara}
\par
The anomaly itself given in eq.(\ref{nonono}), corresponds to
$f=G|\vr-\vR|^{-3}(\vr-\vR)$. In this case, because of $f(\vr)\rar
0$ at infinity, we can evaluate the surface integral in
eq.(\ref{bushlaba}). We have for any choice of $\m(x)$
compatible with MOND, and for any $\eta$
 \beq \vg^{\anom}(\vR)={1\over 4\pi}\int{(\m-1)\over |\vr-\vR|^3}
 [\gf-3(\gf\cdot\vn)\vn]\drt-{1\over 3}(1-\m_g)\vg_g,
\eeqno{neratabar} where $\vn$ is a unit vector in the direction of
$\vr-\vR$.

Another version, which might also be useful, is gotten when writing
expression (\ref{phantomi}) for $\rp$ as
 \beq \rp={1\over 4\pi G}\div[\gf-(\m/\m_g)\gf]+[(1/\m_g)-1]\r.
 \eeqno{phantomus}
This now gives
 \beq F=(1/\m_g-1)F^{\r}+{1\over 4\pi
G}\int(\m/\m_g-1)(\gf\cdot\grad)f(\vr)\drt-{1\over 4\pi
G}\int(\m/\m_g-1)f(\vr) \gf\cdot\vds, \eeqno{nerat}
 where $F^{\r}$
is the quantity $F$ calculated with the true mass, and the
convergence of the surface integral has now been improved.
\subsection{Asymptotic behavior of $\rp$}
\label{A6} The asymptotic behavior of $\rp$ can be derived from the
known behavior of the MOND field there. As derived in Bekenstein \&
Milgrom (1984), the internal potential is, asymptotically (for
$u\gg\eta^{-1/2}R\_M$),
 \beq \f\_{in}\approx -{MG\over \m_g(1+L_g)^{1/2}}\left(x^2+y^2+{z^2\over 1+L_g}\right)^{-1/2}.\eeqno{potat}
 The asymptotic acceleration field is thus
 \beq \vg\_{in}\approx -{MG\over \m_g(1+L_g)^{1/2}}\left(x^2+y^2+{z^2\over 1+L_g}\right)^{-3/2}[x,y,{z\over 1+L_g}],\eeqno{potatita}
 and the phantom density
 $$\rp\approx {-M\over 4\pi\m_g}{L_g\over (1+L_g)^{3/2}}\left(x^2+y^2+{z^2\over
 1+L_g}\right)^{-5/2}\left(x^2+y^2-{2z^2\over 1+L_g} \right)=$$
 \beq = -{M\over 4\pi  \m_g}{L_g\over
 (1+L_g)^{3/2}}u'^{-3}(1-3z'^2/u'^2), \eeqno{mlyul}
where $x'=x$, $y'=y$, $z'=z(1+L_g)^{-1/2}z$,
$u'=(x'^2+y'^2+z'^2)^{1/2}$. We see that $\rp$ decreases
asymptotically as $u^{-3}$.
\par
The $z$ component of the anomalous acceleration on the $z$ axis and the $x$
component on the $x$ axis are respectively

 \beq g\^{\anom}_z(z)\approx -{MG\over z^2}\left[{1\over
\m_g}-1\right], ~~~~
 g\^{\anom}_x(x)\approx
 -{MG\over x^2}\left[{1\over \m_g(1+L_g)^{1/2}}-1\right] \eeqno{jutirw}

 We learn from this that at the same distance along the two axes,
the MOND acceleration on the symmetry axis is larger by a factor
$(1+L_g)^{1/2}$ than that in the perpendicular direction: there is a
stronger pull towards the Sun along the $z$ axis. This is then also
the case for the anomalous acceleration, since the Newtonian one is
spherical. In fact, we see that the anomaly along the $z$ axis is
always attractive towards the sun, asymptotically. This is the
opposite to what a negative value of $q$ gives near the sun.
\section{Properties of the surrogate phantom mass}
\label{B}

\subsection{Its Newtonian field vanishes at the position of the sun}
\label{B3} Here I show that the Newtonian field of the surrogate,
phantom matter vanishes at the position of the sun. I first show, in
general, that the force of a true density $\r$ on its surrogate,
phantom mass vanishes identically. This force is
 \beq \tilde\vF^p=-\int\trp\gf^{\r}\drt={1\over 4\pi
 G}\int(\div\vg^*+4\pi G\r)\gf^{\r}\drt, \eeqno{kutres}
 where $\gf^{\r}$ is the Newtonian field of $\r$ alone, and the
integral is over all space. The second term vanishes, because it
constitutes the Newtonian force of $\r$ on itself. The $i$th
component of $\tilde\vF^p$ can then be written as
 \beq \tilde F^p_i={1\over 4\pi
 G}\int\div\vg^*\f^{\r}_{,i}\drt=-{1\over 4\pi
 G}\int(\n\f\^N_{,k})_{,k}\f^{\r}_{,i}\drt=-{1\over 4\pi
 G}\int[(\n\f\^N_{,k}\f^{\r}_{,i})_{,k}-\n\f\^N_{,k}\f^{\r}_{,i,k}]\drt. \eeqno{kutresaq}
 Here, $\f\^N$ is the potential of the full Newtonian field--that of
 $\r$ plus the background field: $\vgN=-\gf\^N$, and $\n$ is a function
of $|\vgN|/\az$. (A subscript $,i$ is the derivative with respect to
the $i$th coordinate.) In the second term in the integrand we can
replace $\f^{\r}_{,i,k}$ with  $\f\^N_{,i,k}$, because
$\f^{\r}_{,i}$ and $\f\^N_{,i}$ differ by the constant background
acceleration. We then write this term, now in the form
$\n\f\^N_{,k}\f\^N_{,k,i}=N_{,i}[(\gf\^N/\az)^2]$, where
$N'(z)\equiv \n(\sqrt{z})/2$. Applying Gauss theorem we then have
 \beq \tilde \vF^p=-{1\over 4\pi G}\int(\n\gf^{\r}\gf\^N\cdot
 \vds-N\vds),
 \eeqno{kutrim}
where the integration is over the surface at infinity. In the first
term $\gf^{\r}$ decreases as $r^{-2}$; so the other factors can be
taken at their constant limiting value at infinity: $\n$ becomes
$\n_g=\n(\eN)$. In the second term, we take the first order
expansion $N[(\gf\^N/\az)^2]\rar N(\eN^2)-\n_g\gf^{\r}\cdot\vg\^N_g$.
The constant, limiting value of $N$ gives a vanishing integral, and
we thus have
 \beq \tilde \vF^p={\n_g\over 4\pi G}\int(\gf^{\r}\vg\^N_g
 \cdot\vds-\gf^{\r}\cdot\vg\^N_g\vds).
 \eeqno{kutrimas}
Taking the $z$ axis in the direction of $-\vg\^N_g$, we have
 \beq \tilde \vF^p=-{\n_g g\^N_g\over 4\pi
 G}\int(\gf^{\r}d\s_z-\f^{\r}_{,z}\vds).
 \eeqno{kutbe}
By applying Gauss's theorem in reverse, writing the right hand side
as a volume integral, we see that the two terms cancel, and thus
$\tilde \vF^p=0$. This also means that the net force of the phantom
matter on the true matter always vanishes. For our special case, it
follows that the Newtonian force of $\trp$ on the Sun vanishes.
However, since the Sun is a point mass, it follows that the
Newtonian acceleration field of $\trp$ vanishes at the position of
the sun.
\subsection{Its column density along the symmetry axis}
\label{B4} It is easy to gather that $\trp$ can be gotten from
equations exactly like eqs.(\ref{phantomu})(\ref{phantoma}), with
$\gf$ there replaced by $-\vg^*$, with the same $\U$ function. Its
column density along the symmetry axis is thus identical to that of
$\rp$, and equals $\az G^{-1}\c$ with
$\c=(2\pi)^{-1}\int_0^{\infty}L(x)$.

\subsection{it takes up both signs}
\label{B2} As in the case of $\rp$, $\trp\propto
\grad|\vgN|\cdot\vgN$, and for similar reasons this takes up both
signs on any surface of constant $|\vgN|$ surrounding the critical
point, and excluding true matter. This can also be seen explicitly
in expression (\ref{hittra}) for $\hrp$, which is a good
approximation for $\trp$ around the critical point.
\subsection{The total mass}
\label{B1}
 Use eq.(\ref{hutaf}) for $\trp$ to
write
 \beq \trp=
 -{1\over 4\pi G}\div[(\n-1)\vgN].    \eeqno{hutder} Applying Gauss's
theorem to eq.(\ref{hutaf}) we have
 \beq \tilde M_p=-{1\over 4\pi G}\int(\n-1)\vgN\cdot\vds,  \eeqno{mibba}
on the surface at infinity, where we can write $\vgN\rar
\vg_g\^N+\vg^{\r}$. Here $\vg_g\^N$ is the Newtonian external field,
and $\vg^{\rho}=-M G\vu u^{-3}$ is the asymptotic Newtonian field of
the true mass alone. Expanding around $\vg_g\^N$, we get
 \beq \tilde M_p/M=\n(\eN)-1+\eN\n'(\eN)/3,  \eeqno{hutqa}
where $\eN=\m(\eta)\eta$ is the Newtonian galactic field. Using the
relations $\n(\eN)=1/\m(\eta)$ and
$\n'(\eN)=-\eta^{-1}\m(\eta)^{-2}L(\eta)[1+L(\eta)]^{-1}$ we have
 \beq \tilde M_p/M={3+2L_g\over 3\m_g(1+L_g)}-1.
   \eeqno{hutqasa}
   Since $0\le L_g\le 1$ this gives very similar values to those of
   the exact expression (\ref{asym}). In the MOND regime
   $L_g\approx 1$, $\m_g\ll 1$ it gives $5/6\m$ compared with
   $\pi/4 \m$ for $M_p$.

\subsection{integrals weighted by $\trp$}
\label{B6}
\par
We shall also need similar expression to eq.(\ref{bushlaba}) when
$\trp$ is used instead of $\rp$. Repeating the arguments in \ref{A5}
using expression (\ref{hutder}), we see that we get an expression
similar to eq.(\ref{bushlaba}) with $\gf$ replaced by $-\vg^*$. For
example, the surrogate anomaly is
 \beq \tilde\vg^{\anom}(\vR)={1\over 4\pi}\int{(\n-1)\over
|\vr-\vR|^3}[\vgN-3(\vgN\cdot\vn)\vn]\drt-{1\over
3}(\n_g-1)\vg\^N_g. \eeqno{neratfar}
\par
The surrogate quadrupole moment for the limiting form of $\m$, which I shall need below, is
 \beq \tilde Q_{zz}=-{1\over 4\pi G}\int(\n-1)(\vr\cdot\vgN
-3zg\^N_{z})\drt. \eeqno{neratar}
\subsection{The asymptotic behavior of $\trp$}
\label{B7} Using the algebraic relation (\ref{algebraic}) it is
readily seen that the asymptotic behavior of $\vg^*$ is
 \beq \vg^*-\vg_g\approx -{MG\over \m_g u^3}\left(x,y,{z\over 1+L_g}\right).\eeqno{puzata}
The asymptotic behavior of $\trp$ derived from eq.(\ref{puzata}) is
 \beq \trp\approx -{M\over 4\pi  \m_g}{L_g\over
 1+L_g}u^{-3}(1-3z^2/u^2) \eeqno{muopl}
[$u=(x^2+y^2+z^2)^{1/2}$]. We see a similar normalization and radial
dependence as in $\rp$, but a different angular dependence.

\subsection{Delineation of $\tP$ for the limiting $\m$ and $\eN>1$}
\label{B5}
 The region $\tP$ is defined by $|\vg^*|<\az$, or
equivalently by $|\vgN|<\az$. The Newtonian acceleration at position
$\vu$ relative to the Sun is given in eq.(\ref{mugter}). The ray
from the Sun making an angle $\t$ with the $z$ axis cuts $\tP$ at
 \beq u_z^{\pm}=\left[{\eN \cos\t\pm(1-\eN^2 \sin^2\t)^{1/2}\over
 (\eN^2-1)\cos^{-2}\t}\right]^{1/2},  \eeqno{lpitr}
    which give the boundaries of $\tP$. The viewing
angle of $\tP$ from the Sun occurs where $u_z^+=u_z^-$, for which
$\sin\t=1/\eN$. $\tP$ intersects the $z$ axis at $u_z^{\pm}=(\eN\mp
1)^{-1/2}$. The critical point is at $\eN^{-1/2}$ (all in units of
$R\_M$).  Thus, the linear size of $\tP$ decreases as $\eN^{-3/2}$
for large $\eN$, while the distance of $\tP$ from the Sun decreases
only as $\eN^{-1/2}$. This will become important below.
\par
For the special case $\eN=1$, $\tP$ extends everywhere to the
positive $z$ side of the surface
$z_-(r)=\{[(r^4+2)^{1/2}-r^2]/2\}^{1/2}$, with the critical point at
$z=1$.

\section{Further approximations for the limiting form of $\m$}
\label{C} Here I derive an approximate, closed expression for the
dimensionless, surrogate anomaly $\tilde q(\eta)$ for the limiting
form of $\m$. It can be used to check numerical evaluations of the
anomaly, and it is also a closed expression for some sort of a lower
limit on the anomaly for $\eta>1$. The approximation is based on the
observation that the linear size of $\tP$ becomes increasingly small
relative to its distance from the sun, as $\eN$ increases (see
\ref{B5}). In this limit we can use for $\vgN$ in $\tP$ its linear
expansion around the critical point, where $\vgN$ vanishes. This
means writing
 \beq \vgN\approx -{\eN^{3/2}\az\over R\_M}(\vr-3z\ve_z),\eeqno{mugef}
instead of the exact expression (\ref{mugter}) ($\vr$ is now measured
from the critical point, unlike $\vu$, which is measured from the
sun). With this approximation, the region $\tP$ becomes the
interior, $\hP$, of the ellipsoid
 \beq x^2+y^2+4z^2=\ell^2,  \eeqno{ellips}
centered at the critical point, where $\ell=\eN^{-3/2}R\_M$.
\par
The phantom density in this approximation, $\hrp$, is given by
 \beq \hrp=-{1\over 6}\r\_{\odot}\eN^{3/4}
 {x^2+y^2-8z^2\over (x^2+y^2+4z^2)^{5/4}}, \eeqno{hittra}
where $\r\_{\odot}=3\msun/4\pi R\_M^3$; it is axisymmetric, but also
symmetric under reflection in $z$. This means that, unlike $\rp$,
and $\trp$, it does not have a dipole moment. One can check that the
Newtonian field of $\hrp$ does not vanish on the $z$ axis, except at
the origin. In particular, in itself, it does not vanish at the
sun's position, as required. We cannot then use expression
(\ref{mugef}) directly in eq.(\ref{anoor}) to get the desired approximate
expression. In replacing $\trp$ by $\hrp$, the quadrupole moment of
the phantom mass, $Q_{ij}$ (not to be confused with the quadrupole
anomaly), is expected to be well approximated, but we have
completely lost the important contribution of its dipole moment
$\vD$, which is needed to balance the quadrupole and annihilate the
field at the sun's position. The dipole moment derives from the
small $z$ asymmetry that does not affect $Q_{ij}$ much. I thus use
the above approximation only for calculating the quadrupole moment.
I imagine that $\hrp$ is slightly distorted so as to attain a dipole
moment. If we know the quadrupole moment we can determine the dipole
moment by requiring that they cancel at the sun, use this to
eliminate the dipole, and express the whole effect in terms of the
quadrupole. To recapitulate, I approximate the anomaly--the
Newtonian field of the phantom mass-by its dipole-plus-quadrupole
contribution, writing for the potential\footnote{$\vD=\int
\rp(\vr)\vr \drt$, and $Q_{ij}=-{1\over
 2}\int\rp(\vr)(r^2\d_{ij}-3r_ir_j)\drt$.
 $Q_{ij}$ is a traceless matrix, and with the
symmetry of the problem it is diagonal, with
$Q_{xx}=Q_{yy}=-Q_{zz}/2$.} \beq \hat\f^{\anom}(\vR)=-{G\over
R^3}\vD\cdot\vR-{G\over R^5}R^iR^jQ_{ij}. \eeqno {field} ($\vR$,
like $\vr$, is measured from the critical point.) Now, to have the
acceleration field of $\hat\f^{\anom}$ vanish at the sun's position
we have to have\footnote{This relation does not need to hold exactly
since higher multipoles may also contribute to the field value at
the sun, but it should hold to the same accuracy as the
dipole-plus-quadrupole approximation.}
 \beq \vD={-3\over 2\tRz}Q_{zz}\ve_z.  \eeqno{nuti}
(Here, $\tRz$ is the distance from the Sun to the critical point of
$\tP$; $\tRz=\eN^{-1/2}R\_M$.) I put this value of $\vD$ back into
the expression for $\hat\f^{\anom}$ to get the desired anomaly in
this approximation, which requires only knowledge of the quadrupole
moment.
 From this we can identify the anomalous quadrupole and find
 \beq \hat q^{\anom}_{xx}=\hat q^{\anom}_{yy}=-{1\over 2}\hat q^{\anom}_{zz}=
 {3G\over 2R_0^5}Q_{zz}\equiv -{\az\over 2R\_M}\hat q(\eta).
 \eeqno{anom}
\par
Using expression (\ref{mugef}) for $\vgN$ in eq.(\ref{neratar})--the
formula for the quadrupole moment--then gives straightforwardly
 \beq Q_{zz}\approx{\az R\_M^4\over 90 \eN^{6}G}.
\eeqno{qzzi}
 The scaling with this particular power of $\eN$ is
understood as follows: For large $\eN$, the linear size, $d$, of
$\tP$ is $\propto R\_M\eN^{-3/2}$ in all directions . The
characteristic surface column density is always $\S\propto \az/G$;
the quadrupole moment, which scales as $\S d^4$ must then be
$Q\propto \az G^{-1}R\_M^4 \eN^{-6}$.
 This gives finally
 \beq \hat q(\eta)=-\eN^{-7/2}/30.
 \eeqno{anomasa}

\par
Note that our expressions (\ref{qzzi})(\ref{anomasa}), and the
approximation behind them, must break down for values of $\eta$
slightly above 1, or lower. We know, for example, that the anomaly
should vanish in the limit $\eN\rar 0$, which relation
(\ref{anomasa}) fails completely to account for.

\clearpage

\end{document}